\documentclass[%
 aip,
 amsmath,amssymb,
 reprint,%
]{revtex4-1}

\usepackage{graphicx}
\usepackage{dcolumn}
\usepackage{bm}

\usepackage[utf8]{inputenc}
\usepackage[T1]{fontenc}
\usepackage{mathptmx}
\usepackage{etoolbox}
\usepackage{makecell}
\newcommand{\etal}{\textit{et al.}}
\usepackage{colortbl}
\usepackage{siunitx}
\usepackage{colortbl}
\usepackage[dvipsnames]{xcolor}
\usepackage{multirow}
\usepackage{relsize} 
\usepackage{xurl}
\usepackage[normalem]{ulem}

\newcommand{\black}{\color{black}}  
\definecolor{revtext}{RGB}{128,0,160}

\makeatletter
\def\@email#1#2{%
 \endgroup
 \patchcmd{\titleblock@produce}
  {\frontmatter@RRAPformat}
  {\frontmatter@RRAPformat{\produce@RRAP{*#1\href{mailto:#2}{#2}}}\frontmatter@RRAPformat}
  {}{}
}%
\makeatother
\AtBeginDocument{\def\selectlanguage#1{}}

\begin{document}

\preprint{AIP/123-QED}

\title[BO of LA-CNT targets for 800 nm LWFA in the bubble regime]{Bayesian optimization of longitudinally aligned CNT targets for 800 nm laser wakefield acceleration in the bubble regime}
\author{B.S. Nunes}
\affiliation{Nuclear and Energy Research Institute, IPEN–CNEN, São Paulo 05508-000, Brazil}
\email{brunosnunes@usp.br, abonatto@ufcspa.edu.br}

\author{P. Mart\'in-Luna}
\affiliation{Instituto de F\'isica Corpuscular (IFIC), Universitat de Val\`encia - Consejo Superior de Investigaciones Cient\'{\i}ficas, 46980 Paterna, Spain}
\affiliation{Laboratory of Earth Observation, Image Processing Laboratory, University of Valencia, 46980 Paterna, Valencia, Spain}

\author{J. Rodr\'iguez-P\'erez}
\affiliation{ELI Beamlines Facility, The Extreme Light Infrastructure ERIC, Doln\'i B\v{r}e\v{z}any, Czech Republic}
\affiliation{Departamento de F\'isica Aplicada y Electromagnetismo, Universidad de Val\`encia, Burjassot, 46100, Spain}

\author{A. Sytov}
\affiliation{INFN Sezione di Ferrara, Via Saragat 1, Ferrara, 44122, Italy}

\author{B. Lei}
\affiliation{Department of Physics, University of Liverpool, Liverpool, L69 7ZE, United Kingdom}

\author{G. Xia}
\affiliation{Department of Physics and Astronomy, University of Manchester, Manchester, M13 9PL, United Kingdom}
\affiliation{Cockcroft Institute, Keckwick Lane, Daresbury, Warrington WA4 4AD, UK}

\author{J. Resta-López}
\affiliation{Departamento de F\'isica Aplicada y Electromagnetismo, Universidad de Val\`encia, Burjassot, 46100, Spain}

\author{M.S. Alva-Sánchez}
\affiliation{Graduate Program in Information Technology and Healthcare Management, Federal University of Health Sciences of Porto Alegre (UFCSPA), Porto Alegre 90050-170, Brazil}

\author{N.D. Vieira Jr.}
\affiliation{Nuclear and Energy Research Institute, IPEN–CNEN, São Paulo 05508-000, Brazil}

\author{R.E. Samad}
 \affiliation{Nuclear and Energy Research Institute, IPEN–CNEN, São Paulo 05508-000, Brazil}
 
\author{A. Bonatto}
\affiliation{Graduate Program in Information Technology and Healthcare Management, Federal University of Health Sciences of Porto Alegre (UFCSPA), Porto Alegre 90050-170, Brazil}

\date{\today}

\begin{abstract}
Laser wakefield acceleration (LWFA) in solid-density plasmas can reach accelerating gradients in the \si{\tera\eV/\meter} range, well above those typical of gaseous media, provided that the driver is matched to the plasma. In nanostructured solids, this matching can be achieved by controlling the spacing between the constituent nanostructures to reduce the effective density, enabling infrared pulses to excite wakefields in the bubble regime. Following numerical demonstrations with carbon nanotube (CNT) bundles distributed on concentric cylindrical shells, a simpler and experimentally more accessible longitudinally aligned CNT (LA-CNT) target is investigated here, with the bundles on a triangular lattice within a square cross section and an optional central aperture. Fully three-dimensional particle-in-cell simulations are coupled with Bayesian optimization to vary the effective density, target length, bundle radius and aperture radius for a fixed 100~\si{\milli\joule}, few-cycle, 800~\si{\nm} laser pulse. The laser-to-beam energy conversion efficiency is maximized for extracted electrons whose forward longitudinal momenta correspond to kinetic energies above 20~\si{\MeV}. With the aperture radius constrained to at least 0.5~\si{\micro\meter}, the best target operates in the bubble regime and reaches an efficiency of 24.6\%, delivering a 673~\si{\pico\coulomb} bunch with a mean energy of 36.6~\si{\MeV} and a quasi-monoenergetic peak near 46~\si{\MeV}, against 19\% for the concentric-shell target simulated under identical numerical conditions. When the aperture is allowed to vanish, the optimization converges to denser and shorter targets, the efficiency peaks at 25.8\%, and the best configurations deliver a higher charge at the cost of a lower peak energy and an approximately 50\% larger divergence. The aperture-free targets among them, which require no bundle removal, retain a comparable mean efficiency of 24.1\% with the same trade-off. The optimized targets are also robust to fabrication imperfections. Random bundle displacements that preserve the effective density leave charge and efficiency essentially unchanged and produce only modest spectral shifts. These results highlight the role of effective-density control in achieving efficient infrared-driven acceleration in LA-CNT targets, even without a central aperture. By eliminating the need to pattern and align a micrometre-scale opening, aperture-free targets may facilitate proof-of-principle experiments on solid-state LWFA and offer a simpler option for applications that can tolerate their reduced beam quality.
\par
\end{abstract}

\maketitle

\section{Introduction}\label{sec1}
In laser wakefield acceleration (LWFA), accelerating fields far beyond those available in conventional radio-frequency structures are generated through the collective response of plasma electrons to an intense, ultrashort laser pulse~\cite{Tajima1979,Esarey2009}. In a typical configuration, the pulse propagates through a gaseous medium, displaces the plasma electrons and excites a trailing plasma wave that sustains gradients of tens of \si{\giga\eV/\meter}, in which electrons can be trapped and accelerated at the rear of the pulse. Compact sources of energetic electron bunches have been developed with this mechanism~\cite{Mangles2004}, and applications have been investigated in medical radioisotope production~\cite{Vieira2021,Nunes2025,Nunes2026b}, radiotherapy~\cite{Labate2020,Santos2025}, industry~\cite{Assmann2020}, and high-energy physics~\cite{Schroeder2010,Schroeder2016}.

Still shorter acceleration distances can be envisaged by extending LWFA to solid-state plasmas. The characteristic plasma-field scale, $E_0=m_e c\omega_p/e$, is proportional to $\sqrt{n_e}$, where $n_e$ is the free-electron density and $\omega_p$ the plasma frequency~\cite{Esarey2009}. Solid-density plasmas ($10^{22}$--$10^{24}~\si{\per\centi\meter\cubed}$) are several orders of magnitude denser than gaseous media and could therefore sustain accelerating gradients in the \si{\tera\eV/\meter} range, two to three orders of magnitude above those of gas-based LWFA~\cite{Tajima1987,Zhang2016}. Solid targets also offer practical advantages: they are mechanically self-supporting, can be engineered with nanometre precision, and require no gas handling or differential pumping, so that the plasma source itself becomes a microscopic, potentially mass-produced component. Such accelerators, if realized, would enable ultracompact sources of relativistic electrons and secondary radiation for the applications listed above, and the \si{\tera\eV/\meter} fields would provide a new testbed for strong-field and high-energy-density physics. Nanostructured solids also support collective modes that have no counterpart in gaseous plasmas, such as surface plasmons on the inner wall of carbon nanotube (CNT) channels and solid microtubes. The excitation of these modes, either by optical lasers~\cite{Lei2025PPCF,Lei2026PRL} or by dense electron beams~\cite{Lei2025NJP}, has been proposed as a wakefield acceleration mechanism, with predicted gradients of hundreds of \si{\tera\eV/\meter}, and laser-excited surface plasmon polaritons have also been suggested as a route toward compact sources of coherent synchrotron radiation~\cite{Lei2025PRL}.

The challenges, however, are equally significant. At solid density, the plasma wavelength and period shrink to tens of nanometres and sub-femtosecond timescales, respectively. Driving a wakefield in a bulk solid would therefore require ultrashort X-ray pulses of extreme intensity, which are not yet available, whereas the propagation of optical pulses, and hence sustained wakefield excitation, is hindered by the strongly overdense character of bulk solids. The dephasing and depletion lengths are also reduced to the micrometre scale and, since the target is expected to be damaged or destroyed, it must be replaced and realigned between shots. Finally, nanostructures with a prescribed geometry must be fabricated with sufficient reproducibility.

The idea of exploiting the periodic structure of a solid to sustain accelerating wakefields dates back to the crystal X-ray accelerator proposed by Tajima and Cavenago~\cite{Tajima1987}, and the first particle-in-cell (PIC) studies of nanostructured targets followed the same logic, considering wakefields driven by a 1~\si{\nm} X-ray pulse in CNTs modelled either as homogeneous solid-density cylinders or as hollow plasma channels~\cite{Zhang2016}. In these works, the nanostructure essentially provides a solid-density plasma, the driver wavelength has to be scaled down to match it, and arrays of individual CNT bundles are not resolved.

A different route, pursued within the NanoAc collaboration~\cite{NanoAc2022,NanoAc2024,NanoAc2025,NanoAc2026}, is to use the voids of the nanostructure to tune the density experienced by the driver, so that wakefields can be excited in solid-state targets with the near-infrared, high-power lasers already available in many laboratories. By increasing the spacing between graphene layers or CNT bundles, the filling fraction and the spatially averaged density are reduced while the local material density is retained. Once the driver ionizes the nanostructure, the displaced electrons sustain a collective wakefield. As long as the period of the structure is small compared with the plasma wavelength and the driver spot size, the large-scale response is that of a homogeneous plasma at the spatially averaged, or effective, density. Analytical results derived for homogeneous plasmas can therefore be applied to the nanostructured medium, with the effective free-electron density in place of $n_e$.

For concentric tubes with walls of width $w$ separated by gaps $g$, modelled as a pre-ionized plasma~\cite{Bonatto2023}, the effective free-electron density is the electron density of the walls divided by $\kappa=(w+g)/w$; for a large number of tubes, $1/\kappa$ tends to the fraction of the volume occupied by the walls, that is, to the filling fraction $f$. In this work, the target consists of longitudinally aligned CNT (LA-CNT) bundles, i.e., bundles parallel to the laser propagation direction, modelled as homogeneous cylinders of radius $r_b$ and bundle-averaged carbon number density $n_b$. The filling fraction is the ratio of the volume occupied by the bundles to the total target volume, which for parallel bundles of equal length reduces to the fraction of the transverse cross section that they cover. Since the target is initially neutral and is ionized by the laser itself, the effective carbon number density, $n_{\mathrm{eff}}=f\,n_b$, is fixed by the geometry, whereas the effective free-electron density, $n_{e,\mathrm{eff}}$, depends on the ionization state of the carbon ions and cannot exceed $6\,n_{\mathrm{eff}}$, the value reached for fully ionized carbon.

Reducing the filling fraction at a given ionization state lowers $n_{e,\mathrm{eff}}$ and increases both the characteristic plasma wavelength, $\lambda_{p,\mathrm{eff}}=2\pi c/\omega_{p,\mathrm{eff}}\propto n_{e,\mathrm{eff}}^{-1/2}$, and the corresponding oscillation period~\cite{Esarey2009,Bonatto2023}. A laser pulse of longer wavelength and duration, in the ultraviolet or the infrared, can then be accommodated within a fraction of this enlarged wakefield wavelength, which, together with the intensity and focusing required for electron cavitation, is the condition for driving wakefields in the nonlinear or bubble regime~\cite{Esarey2009,Bontoiu2025}. Solid targets can thus be designed for a given infrared driver by adjusting the filling fraction of the nanostructure. \black

This approach has been developed and tested in a sequence of numerical studies. A first step was a 2D Cartesian EPOCH~\cite{Arber2015} study of beam-driven wakefields in an array of alternating plasma layers and empty channels representing a CNT array, reported at IPAC 2018~\cite{RestaLopez2018}. The effective-density description adopted in this work was first presented at the ``Applications of Crystals and Nanotubes for Acceleration and Manipulation (ACN2020)'' workshop~\cite{Bonatto2020,ACN2020Summary}. In that contribution, a simple analytical model and PIC simulations were used to show that, under suitable conditions, wakefields excited in periodically structured plasmas, such as those representing CNT arrays, can be described by homogeneous-plasma theory evaluated at an effective density. Parameter scans were performed for beam-driven wakefields, and preliminary results were also presented for X-ray-laser drivers. The notion of an effective electron density perceived by the laser, set by the spacing between the nanotubes, was later adopted by Nicks \etal~\cite{Nicks2021} in a high-density LWFA study based on 1D and 2D EPOCH simulations, in which electron acceleration was evaluated for CNT bundles aligned parallel to the laser propagation direction and for CNT arrays oriented perpendicular to it. The systematic formulation of the approach for beam-driven wakefields, based on the concentric-tube model described earlier, followed in 2023~\cite{Bonatto2023}, using 2D axisymmetric (quasi-3D) simulations with the Fourier--Bessel PIC code FBPIC~\cite{Lehe2016}. Effective-density control is now one of the pillars of target design within the NanoAc collaboration~\cite{Bonatto2026}.


Laser-driven acceleration in spaced nanostructures, the route followed here, was then demonstrated numerically for a stack of equally spaced graphene layers with a linearly polarized, 100~\si{\nm}, 3~\si{\femto\second} (nine-cycle) pulse with $a_0=2.7$ and 8~\si{\milli\joule} of energy, modelled with 2D Cartesian PIC simulations using the PIConGPU code~\cite{Burau2010,Bontoiu2023}. Electron self-injection occurred through a new mechanism associated with edge plasma oscillations, named catapult injection, followed by acceleration at an average gradient of approximately 4.8~\si{\tera\eV/\meter}, which produced a 2.5~\si{\pico\coulomb} bunch with a mean energy of about 7~\si{\MeV} and a 7\% energy spread. These results confirmed that interlayer gaps make a solid nanostructure accessible to a driver much longer than the X-ray wavelengths required by bulk solids, although a 100~\si{\nm} source with the required intensity is still not available.

The modelling was then extended to 3D Cartesian geometry and to the 800~\si{\nm} wavelength of Ti:sapphire lasers by Bon\c{t}oiu \etal~\cite{Bontoiu2025}, who considered a three-cycle, circularly polarized pulse with a peak power of 35.5~\si{\tera\watt} interacting with a target of LA-CNT bundles. Each bundle was represented by a homogeneous carbon cylinder whose density was averaged over the unresolved tubes and voids within it, so that the effective-density concept was applied at the bundle level as well. In the transverse plane, their positions were prescribed on virtual concentric cylindrical shells surrounding a central aperture of radius $r_{\mathrm{in}}=0.8~\si{\micro\meter}$, with the number of bundles on each shell chosen to approximate $n_{\mathrm{eff}}=10^{20}~\si{\per\centi\meter\cubed}$ throughout the structured region; both random and regular angular arrangements were considered. With this choice of density and gaps, numerical simulations demonstrated --- for the first time, to the authors' knowledge --- that LWFA driven by an 800~\si{\nm} laser pulse can operate in the bubble regime in a nanostructured solid target. The resulting self-injected bunch carried a sub-nanocoulomb charge and reached a mean energy of about 28~\si{\MeV}, with an average accelerating gradient of 1.9~\si{\tera\eV/\meter}. This target configuration has since been re-simulated at higher resolution with WarpX~\cite{WarpX}, and the accelerated beam, previously defined by selecting the particles inside the bubble, has been redefined as all particles extracted from the target with energies above a given threshold, a criterion closer to what is measured in an experiment~\cite{Bonatto2026}. The influence of the laser polarization on the acceleration dynamics has also been assessed~\cite{Nunes2026}.

While these results represent a major step toward defining the geometry of a nanostructured target for an ultracompact LWFA experiment based on infrared lasers, arranging the bundles in ordered concentric shells poses significant fabrication challenges. This configuration was inherited from the concentric hollow-tube model of Ref.~\cite{Bonatto2023} and, in light of the effective-density picture, may not be necessary, since alternative bundle distributions with the same $n_{\mathrm{eff}}$ are expected to exhibit similar behaviour. In the present study, a simpler LA-CNT target architecture is therefore proposed and investigated within the effective-density approach. The bundles are arranged in staggered rows following a triangular lattice, such that each interior bundle has six equidistant nearest neighbours. The resulting structure has a square transverse cross section with a central aperture, eliminating the need for concentric shells and offering a geometry that may be more straightforward to fabricate.

Fully 3D WarpX PIC simulations~\cite{WarpX} are combined with Bayesian optimization~\cite{Brochu2010}, a method widely applied in laser--plasma acceleration studies~\cite{Shalloo2020,Jalas2021,Dolier2022,Irshad2024,Nunes2025,Santos2025,Nunes2026b}. All parameters of the 800~\si{\nm} laser pulse are kept fixed, while the effective target density, target length, bundle radius, and aperture radius are varied to maximize the laser-to-beam energy conversion efficiency~\cite{Streeter2022,Irshad2024,Nunes2025,Nunes2026b}. Two optimization campaigns are carried out, differing only in whether the central aperture is constrained to a minimum radius or allowed to vanish. In the first campaign, with $r_{\mathrm{in}}\geq0.5~\si{\micro\meter}$, the best configurations operate in the bubble regime and produce spectra with a quasi-monoenergetic peak (QMEP) centred near 46~\si{\MeV} for the optimal target and near 42~\si{\MeV} for the average of the 50 best cases, with a charge of about 670~\si{\pico\coulomb} above 20~\si{\MeV} and a conversion efficiency of 24.6\% for the optimal target.

The robustness of the effective-density description for this geometry is further assessed by introducing Gaussian random displacements to the bundle positions while \(n_{\mathrm{eff}}\) is kept approximately constant. The sensitivity of the acceleration process to positional disorder, representative of potential fabrication imperfections, is thereby investigated. Despite the resulting spatial disorder, the accelerated charge and energy-conversion efficiency are found to remain largely unchanged, although moderate shifts in the high-energy spectral peak are observed. The use of effective density as a design parameter for predicting global acceleration performance is therefore supported within the investigated parameter range, while a residual dependence of spectral features on the local spatial arrangement of the bundles is identified.

Since the first campaign indicated a preference for apertures near the lower bound of the investigated range, the second campaign allowed $r_{\mathrm{in}}$ to reach zero. Aperture-free targets, in which no bundles are removed to form a central channel, yield qualitatively similar spectra and comparable conversion efficiencies, but with the QMEP centred at a lower energy, near 38~\si{\MeV} in the averaged spectrum, a larger beam divergence, and more charge at low energies. These drawbacks come with a practical advantage for experiments in which targets are destroyed on every shot: eliminating the aperture removes the need to fabricate a micrometre-scale channel and to align the laser with it, so that target replenishment may be simplified, albeit at the expense of peak energy and beam quality. The results of both campaigns are compared with those obtained for the concentric-shell target under identical numerical conditions.

This paper is organised as follows. Section~\ref{sec:model} describes the target geometry, the PIC simulation model, and the optimization procedure. Section~\ref{results} presents the results of the two optimization campaigns, the analysis of positional disorder, and the comparison with the concentric-shell target. Finally, the main findings and conclusions are summarized in Sec.~\ref{conclusion}.
\par
\black

\section{\label{sec:model}The Model}

The physical system considered in this study consists of an $800\,\si{\nm}$ laser pulse interacting with an LA-CNT target in vacuum. The laser propagates along the $z$-axis, with its axis coinciding with the target axis at $x=y=0$, while the front face of the target is located at $z=0$. The laser pulse is modelled with the few-cycle pulse profile of Caron and Potvliege~\cite{Caron1999}, which reduces to the standard Gaussian pulse when the envelope contains many optical cycles. The laser is linearly polarized along the $y$ direction, and all its parameters, listed in Table~\ref{tab:laser}, are kept fixed throughout the optimization.

As described in Sec.~\ref{sec1}, the LA-CNT target consists of CNT bundles aligned along the laser propagation direction, each modelled as a homogeneous cylinder of initially neutral carbon atoms. Following a previously adopted approach~\cite{Bontoiu2025,Bonatto2026}, a fixed bundle-averaged number density of $n_b = 10^{22}\,\si{\per\cm\cubed}$ is assumed. This density accounts phenomenologically for the internal structure of the CNT bundles, including voids and inter-tube packing, and is kept constant throughout the study. The target occupies a square transverse cross section in the $xy$ plane, extending from $-3\,\si{\um}$ to $3\,\si{\um}$ along both axes, with a longitudinal length $L$ along the laser propagation direction ($z$-axis). For each target configuration, the bundle radius $r_b$ and the nominal effective carbon number density $n_{\mathrm{eff}}$ are prescribed. The corresponding filling fraction is calculated as $f=n_{\mathrm{eff}}/n_b$, and the number of bundles $N_b$ is obtained by rounding $fA/(\pi r_b^2)$ to the nearest integer, where $A$ is the transverse cross-sectional area of the target and $\pi r_b^2$ is the cross-sectional area of a single bundle.

\begin{figure}[!b]
    \centering
    \includegraphics[scale=0.8]{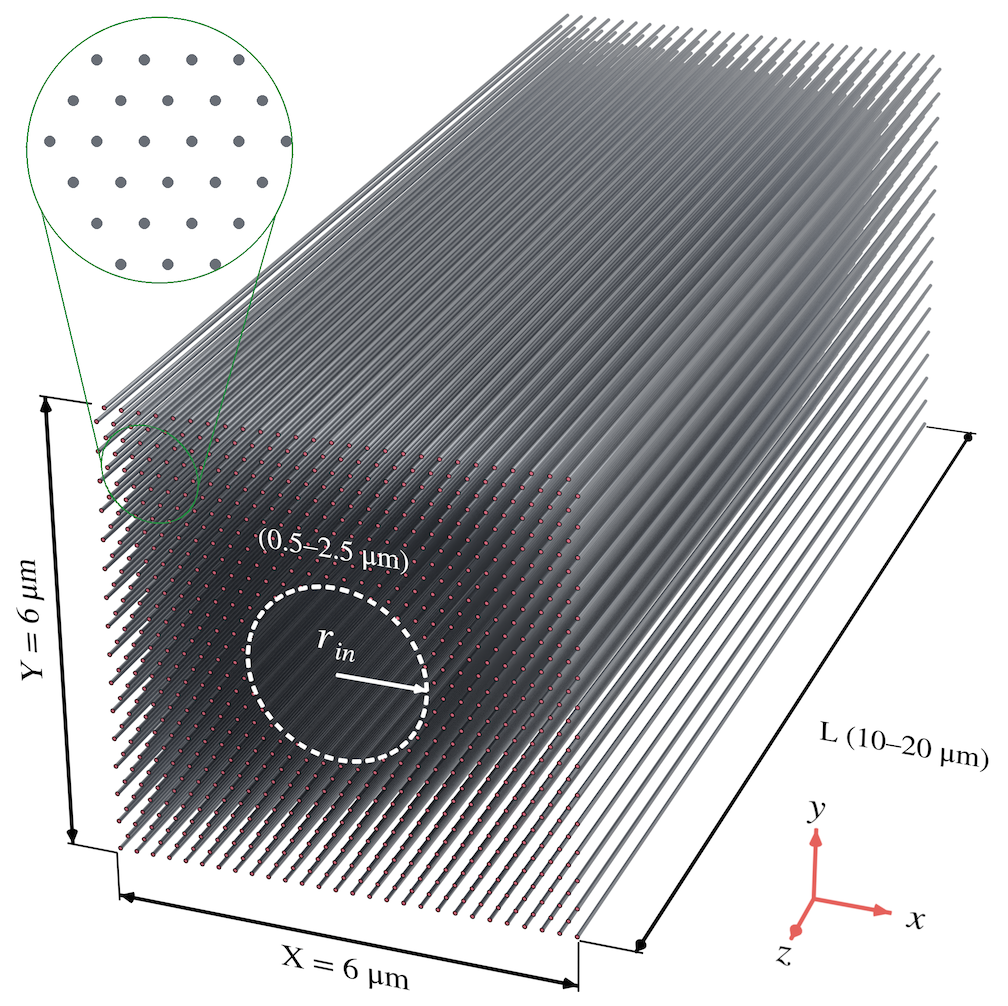}
    \caption{Schematic of the LA-CNT target to be optimized. CNT bundles of density $n_b$ and radius $r_b$ are aligned along the $z$-axis and distributed in regular staggered rows in the transverse $xy$ plane, forming a triangular lattice. The bundle forest has an effective density $n_\mathrm{eff}$ and surrounds a central aperture of radius $r_\mathrm{in}$ along $z$, through which the laser pulse propagates. The target extends from $z=0$ to $z=L$. The green circular inset shows a zoomed view of the bundle arrangement in the $xy$ plane.}
    \label{fig:target}
\end{figure}

Once $N_b$ is determined, the bundles are arranged in regularly staggered rows in the transverse plane, forming a triangular lattice in which the spacing between adjacent rows is $\sqrt{3}/2$ times the in-row spacing. All bundles extend along the $z$-direction over the full target length $L$. Finally, a central aperture of radius $r_{\mathrm{in}}$ is introduced by removing all bundles whose centres satisfy $r<r_{\mathrm{in}}$, where $r=\sqrt{x^2+y^2}$ is the radial distance from the target axis. The resulting aperture provides a channel for laser propagation. The effective density $n_{\mathrm{eff}}$ is defined over the structured region, including the voids between bundles but excluding the central aperture. Introducing the aperture therefore changes the extent of this region without changing its prescribed effective density. The target geometry is illustrated in Fig.~\ref{fig:target}, with the green circular inset showing a magnified view of the bundle arrangement in the transverse $xy$ plane.

\begin{table}[!b]
\caption{\label{tab:laser} Laser parameters.}
\centering
\begin{tabular}{@{}cccl}\hline
\multirow{2}{*}{\makecell[c]{Parameter}} & \multirow{2}{*}{\makecell[c]{Value}} & \multirow{2}{*}{\makecell[c]{Unit}} &\multirow{2}{*}{\makecell[c]{Description}}\\
&&\\
\hline\\[-2ex]
$a_0$ &  21.6 & --- & strength parameter  \\[1.0ex]
$\lambda_0$ & 800 & \si{\nm} & wavelength  \\[1.0ex]
$w_0$ & 1.5 & \si{\um} & waist\\[1.0ex]
 $\tau$ & 2.67 & \si{\femto s} & pulse duration (FWHM) in intensity\\[1.0ex]
 $z_f$ & 0 & \si{\um} & laser focus position\\[1.0ex]
$I_0$ & $10^{21}$ & \si{\watt\per\cm\squared} & peak intensity\\[1.0ex]
$E_\ell$ & 100 & \si{\milli\joule} & pulse energy\\[1.0ex]
\hline
\end{tabular}
\end{table}


\subsection{\label{sec:pic}PIC simulations}

Fully 3D Cartesian PIC simulations were performed with WarpX v.~25.10~\cite{WarpX} on the SDumont supercomputer, using 16 NVIDIA V100 GPUs with 32~GB of memory each. The computational domain comprised approximately $1.1\times10^{9}$ cells of $6.8 \times 6.8 \times 30.3\,\si{\nano\meter}$, initialized with eight macroparticles per cell. The bundles were initialized as neutral carbon with the bundle-averaged density $n_b$ defined in Sec.~\ref{sec:model}; the lower bound of $r_b$ in the optimization, $10\,\si{\nano\meter}$, corresponds to a minimum bundle diameter of $20\,\si{\nano\meter}$, with its cross section discretized into approximately seven mesh cells at this resolution.

Maxwell's equations were solved with the finite-difference time-domain Yee scheme on a staggered grid~\cite{Yee1966}, with the time step set to 0.999 of the Courant--Friedrichs--Lewy limit, and the macroparticles were advanced with the Boris pusher~\cite{Boris1970}. Currents were deposited with the charge-conserving Esirkepov scheme~\cite{Esirkepov2001} using third-order particle shape factors, and perfectly matched layers were applied at all boundaries of the domain. Ionization of the carbon atoms by the laser field was modelled with the Ammosov--Delone--Krainov (ADK) tunnelling rate~\cite{Ammosov1986}, starting from neutral atoms; collisions were not included. To ensure reproducibility and to keep stochastic fluctuations out of the optimization, the same arbitrarily chosen random seed was used in all simulations. The complete set of simulation parameters is summarized in Table~\ref{tab:PIC}.

The accelerated beam is defined as the set of electrons that have exited the target with a forward longitudinal momentum corresponding to a kinetic energy above 20~\si{\MeV}; this experimentally motivated definition follows Ref.~\cite{Bonatto2026}. The beam charge and charge-weighted mean kinetic energy are denoted by $Q_{\mathrm{sel}}$ and $\bar{E}_{\mathrm{sel}}$, respectively. The total kinetic energy of the selected electrons is given by $E_b=(Q_{\mathrm{sel}}/e)\bar{E}_{\mathrm{sel}}$, where $e$ is the elementary charge. Besides defining the electron population used in the objective function (Sec.~\ref{sec:bo}), this choice introduces an implicit beam-quality criterion by favouring the conversion of laser energy into longitudinal electron motion, potentially reducing beam divergence.
The same selection criterion and PIC parameters, including the spatial resolution, were used in Refs.~\cite{Nunes2026,Bonatto2026} to re-simulate the concentric-shell target of Ref.~\cite{Bontoiu2025}, allowing a direct comparison in Sec.~\ref{results}.

\begin{table}[!t]
\caption{\label{tab:PIC} PIC parameters.}
\centering
\begin{tabular}{@{}cccl}
\hline
\multirow{2}{*}{\makecell[c]{Parameter}} & \multirow{2}{*}{\makecell[c]{Value}} & \multirow{2}{*}{\makecell[c]{Unit}} &\multirow{2}{*}{\makecell[c]{Description}}\\
&&\\
\hline\\[-2ex]
\multirow{2}{*}{$z_{min}$} &  \multirow{2}{*}{-3.5} & \multirow{2}{*}{\si{\um}} &
\multirow{2}{*}{\makecell[l]{initial boundary of the longitudinal\\simulation domain}} \\
&  & & \\[1.0ex]
 \multirow{2}{*}{$z_{max}$} & \multirow{2}{*}{27.5} & \multirow{2}{*}{\si{\um}} &  \multirow{2}{*}{\makecell[l]{final boundary of the longitudinal\\simulation domain}}\\
  &&&\\[1.0ex]
 \multirow{2}{*}{$(x,y)_{min}$} &  \multirow{2}{*}{-3.5} & \multirow{2}{*}{\si{\um}} &
 \multirow{2}{*}{\makecell[l]{initial boundary of the transverse\\simulation domain}} \\
  &  & & \\[1.0ex]
 \multirow{2}{*}{$(x,y)_{max}$} & \multirow{2}{*}{3.5} & \multirow{2}{*}{\si{\um}} &  \multirow{2}{*}{\makecell[l]{final boundary of the transverse\\simulation domain}}\\
  &&&\\[1.0ex]
 $N_{x,y,z}$ & 1024 &  --- & Number of mesh cells in each direction\\[1.0ex]
  \multirow{2}{*}{$CFL$} & \multirow{2}{*}{0.999} & \multirow{2}{*}{ ---} &  \multirow{2}{*}{\makecell[l]{ratio between the simulation timestep\\and the Courant-Friedrichs-Lewy limit}}\\
  &&&\\[1.0ex]
 $N_{p}$ & 2 & --- & particles per cell along each direction\\[1.0ex]
 \hline
\end{tabular}
\end{table}

\subsection{\label{sec:bo}Bayesian Optimization}

The quantity maximized in this work is the laser-to-beam energy conversion efficiency~\cite{Streeter2022,Irshad2024,Nunes2025,Nunes2026b}, $\eta \equiv E_b/E_\ell$, where $E_\ell$ is the laser pulse energy; throughout the text, $\eta$ is reported as a percentage. The input parameters are the four target quantities defined in Sec.~\ref{sec:model}, $n_{\mathrm{eff}}$, $r_b$, $r_{\mathrm{in}}$, and $L$, varied within the ranges listed in Table~\ref{tab:input}, while all laser parameters are kept fixed.

Bayesian optimization~\cite{Brochu2010} was adopted because it locates the optimum of an expensive objective function with far fewer evaluations than a parameter scan would require, an essential feature when each evaluation is a 3D PIC simulation. It was implemented with the Optimas library, v.~0.7.1~\cite{Pousa2023,Hudson2022}. Starting from an initial set of 10 simulations with input parameters randomly sampled within their ranges, the algorithm builds a surrogate model of the objective function by Gaussian process regression~\cite{Rasmussen2005}. At each subsequent iteration, the next set of input parameters is chosen with the Noisy Expected Improvement (qNEI) acquisition function~\cite{Balandat2019}, which balances the exploration of the parameter space against the exploitation of the regions where $\eta$ is highest. Although the simulations are deterministic for a fixed seed, the noisy variant of the acquisition function was adopted because it is the default in Optimas and is robust to residual numerical noise in the objective. The corresponding PIC simulation is then run and the resulting value of $\eta$ is used to update the surrogate model, while the other beam quantities analysed in Sec.~\ref{results}, namely the selected charge, mean energy, emittance, and divergence, are extracted from the same simulation. The number of iterations performed in each campaign is given in Sec.~\ref{results}.

\begin{table}[!t]
\caption{\label{tab:input}Input parameters to be optimized.}
\centering
\begin{tabular}{@{}cccl}
\hline
\multirow{2}{*}{\makecell[c]{Parameter}} & \multirow{2}{*}{\makecell[c]{Range}} & \multirow{2}{*}{\makecell[c]{Unit}} &\multirow{2}{*}{\makecell[c]{Description}}\\
&&\\
\hline\\[-2ex]
\multirow{2}{*}{$n_\mathrm{eff}$} &
\multirow{2}{*}{$[0.35\;,\,3.5]$} &
\multirow{2}{*}{$10^{20}$ \si{\per\cm\cubed}} &
\multirow{2}{*}{\makecell[l]{effective carbon number\\density of the target}}\\[1ex]
  &&&\\
        \multirow{1}{*}{$L$} &  \multirow{1}{*}{$[10 \;,\, 20]$}& \multirow{1}{*}{\si{\um}}  & \multirow{1}{*}{\makecell[l]{length of the target}}\\[1ex]

     $r_{b}$ & $ [10 \;,\, 50] $&$ \si{\nm}$ & radius of the CNT bundles \\[1ex]
        $r_\mathrm{in}$ & $[0.5 \;,\, 2.5] $&$\si{\um}$ & central hollow of the target\\
          \hline
\end{tabular}
\end{table}
            
\section{Results}\label{results}

\begin{figure}[!t]
    \centering
     \includegraphics[scale=1]{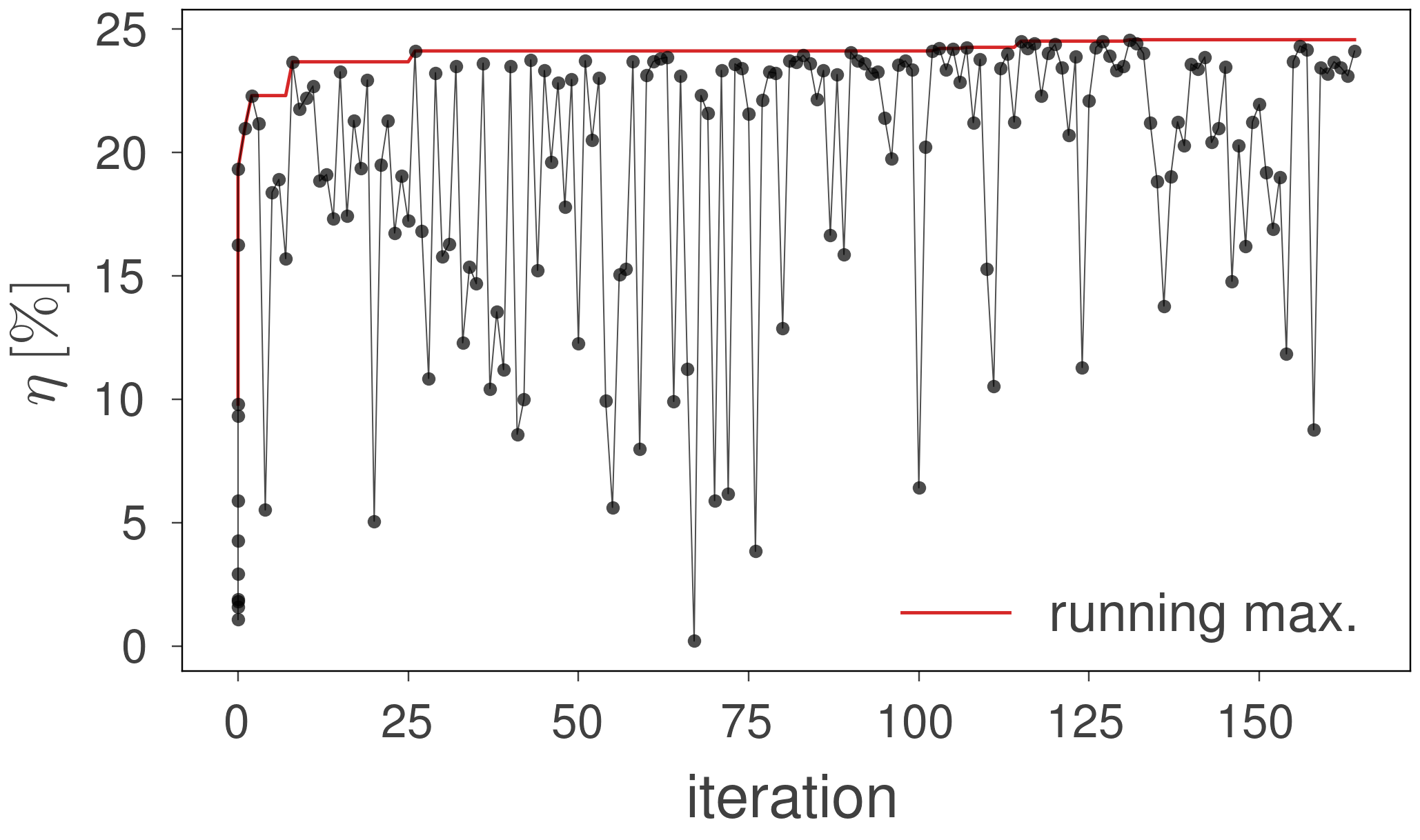}
    \caption{Evolution of the objective function $\eta$ over the course of the Bayesian optimization. Black markers show $\eta$ at each iteration, and the red line indicates the best value obtained up to that point.}
    \label{fig:results_iteration}
\end{figure}

Figure~\ref{fig:results_iteration} shows the evolution of the laser-to-beam energy conversion efficiency, $\eta$, over 167 iterations of the first Bayesian optimization campaign, named $O_1$. Following the 10 initial random evaluations grouped at iteration 0, the algorithm rapidly identifies target configurations yielding high efficiencies. Subsequent evaluations frequently achieve similarly high values of $\eta$, interspersed with lower-efficiency results. This pattern is consistent with the balance between exploration and exploitation provided by the acquisition function described in Sec.~\ref{sec:bo}. The highest efficiency obtained during this campaign is $\eta \approx 24.6\%$. 
To characterize the target configurations associated with high efficiencies and the resulting electron beams, the 50 simulations with the highest $\eta$ values (``50 best'') are selected for analysis. This subset has a mean efficiency of $\eta = (23.9 \pm 0.3)\%$. Hereafter, uncertainties represent the standard deviation across the simulations in each respective set, unless otherwise stated. Shaded bands in the energy spectra represent the mean $\pm 2s$, where $s$ is the standard deviation across the corresponding simulations at each energy bin.

%
\begin{figure*}[!t]
    \centering
    \includegraphics[scale=1]{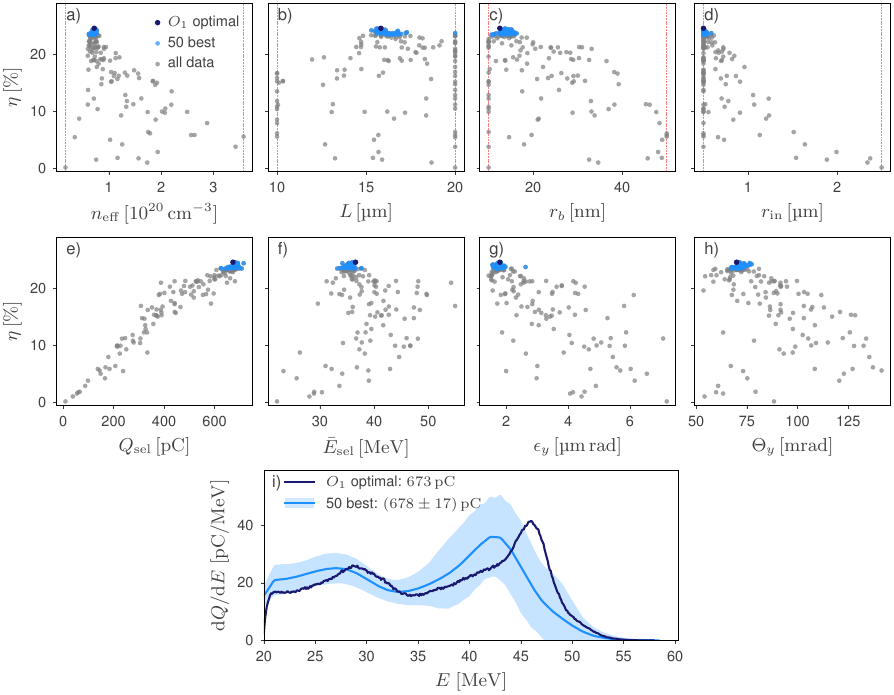}
    \caption{Scatter plots of the Bayesian optimization campaign $O_1$. The dark blue point corresponds to the simulation that achieved the highest $\eta$ value, light blue points indicate the subset comprising the 50 best simulations, and gray points represent the remaining simulations. Input parameters: $\eta$ as a function of (a) the effective target density, (b) the target length, (c) the CNT bundle radius, and (d) the central aperture radius. Output beam parameters---considering only particles with energies above 20~\si{\MeV}: $\eta$ as a function of (e) the beam charge, (f) the average energy, (g) the transverse emittance in the $y$ direction, and (h) the beam divergence in the $y$ direction. (i) Electron energy spectra corresponding to the optimal simulation (dark blue) and to the average of the 50 best (light blue). Shading indicates the mean $\pm 2s$, with $s$ the standard deviation across the 50 best simulations at each energy.}
    \label{fig:results_bayesian}
\end{figure*}

Figures~\ref{fig:results_bayesian}(a)--(d) show the laser-to-beam energy conversion efficiency, $\eta$, as a function of the four target input parameters explored during the first Bayesian optimization campaign, with the red dashed lines indicating the search bounds listed in Table~\ref{tab:input}. Figures~\ref{fig:results_bayesian}(e)--(h) show the corresponding relationships between $\eta$ and the output parameters characterizing the accelerated electron beam obtained from each simulation. Finally, Fig.~\ref{fig:results_bayesian}(i) compares the electron energy spectrum of the optimal configuration with the average spectrum of the 50 best simulations.

The effective carbon number density, $n_{\mathrm{eff}}$, is treated as a continuous input parameter in the Bayesian optimization and used to determine the number of CNT bundles in the target. Since the number of bundles must be an integer, the effective density of the resulting configuration may differ slightly from the value proposed by the optimizer. This discrepancy ranges from 0 to 5\%, and the values shown in Fig.~\ref{fig:results_bayesian}(a) correspond to the actual effective densities used in the simulations. The 50 best configurations cluster around $n_{\mathrm{eff}}=(7.1\pm0.3)\times10^{19}\,\si{\per\cm\cubed}$, indicating a relatively narrow density range associated with high conversion efficiencies. Lower densities, approaching the search bound of $3.5\times10^{19}\,\si{\per\cm\cubed}$, are associated with substantially reduced efficiencies, suggesting less effective electron injection and acceleration. At higher densities, $\eta$ progressively decreases toward the upper bound of $3.5\times10^{20}\,\si{\per\cm\cubed}$. This trend is consistent with the plasma propagation considerations discussed in Ref.~\cite{Bontoiu2025}. Assuming complete ionization of all six carbon electrons, the electron density corresponding to the optimal effective carbon density is approximately $4.3\times10^{20}\,\si{\per\cm\cubed}$, about one-quarter of the critical density $n_c\simeq1.7\times10^{21}\,\si{\per\cm\cubed}$ for an $800\,\si{\nm}$ laser. At the upper bound of the search interval, however, the corresponding fully ionized electron density exceeds the critical density, potentially hindering laser propagation and reducing acceleration efficiency.
\black

Figure~\ref{fig:results_bayesian}(b) shows that high efficiencies are maintained over a relatively broad range of target lengths, $14\,\si{\um} \leq L \leq 20\,\si{\um}$, with several iterations reaching the upper bound, $L = 20\,\si{\um}$. This limit was imposed by GPU memory constraints, as increasing $L$ further while preserving the selected bounds for $n_{\mathrm{eff}}$ and the number of particles per cell would exceed the available memory for some configurations. Regarding the CNT bundle radius, Fig.~\ref{fig:results_bayesian}(c) shows that the 50 best configurations cluster around $r_b = (14 \pm 1)\,\si{\nm}$. In several iterations, the algorithm also explored the lowest admissible value of $r_b$, suggesting a tendency toward smaller radii. However, the lower bound of $10\,\si{\nm}$ was retained to ensure adequate numerical resolution. At this radius, the bundle cross section is resolved by about seven cells (Sec.~\ref{sec:pic}). Smaller radii would be represented by progressively fewer cells, compromising numerical accuracy, and a finer grid was precluded by the available GPU memory. Figure~\ref{fig:results_bayesian}(d) shows convergence toward the minimum allowed inner radius, yielding $r_{\mathrm{in}} = (0.51 \pm 0.02)\,\si{\um}$ for the 50 best configurations. The lower bound corresponds to one-third of the fixed laser waist ($1.5\,\si{\um}$) and was chosen to promote the formation of a wakefield bubble similar to that reported in Ref.~\cite{Bontoiu2025}.

Regarding the output parameters, Figs.~\ref{fig:results_bayesian}(e) and (f) show the selected beam charge and mean electron energy, respectively, considering only macroparticles satisfying the selection criteria. The 50 best configurations yield $Q_{\mathrm{sel}} = (678 \pm 17)\,\si{\pico\coulomb}$ and $\bar{E}_{\mathrm{sel}} = (35.3 \pm 0.8)\,\si{\MeV}$. Although two simulations achieved mean energies above $50\,\si{\MeV}$, their relatively low selected charges resulted in low conversion efficiencies.

Figures~\ref{fig:results_bayesian}(g) and (h) show the RMS normalized emittance and beam divergence along the $y$-axis (parallel to the laser polarization), respectively. Similar results are obtained along the $x$-axis (perpendicular to the polarization) and are therefore not shown. For the 50 best configurations, the mean RMS normalized emittance and beam divergence are $\epsilon_y = (1.8 \pm 0.2)\,\si{\micro\meter\radian}$ and $\Theta_y = (72 \pm 2)\,\si{\milli\radian}$, respectively. Although neither quantity was explicitly included in the objective function, configurations exhibiting relatively low emittance and divergence were obtained among the most efficient solutions. This behavior may be partly attributed to the particle-selection criterion used in the objective function, which favors configurations that efficiently transfer energy to electrons with sufficiently high longitudinal momentum. Since the angular divergence depends on the ratio between transverse and longitudinal momenta, this criterion may indirectly favor the production of more collimated electron beams.

Figure~\ref{fig:results_bayesian}(i) compares the electron energy spectrum of the optimal configuration with the average spectrum of the 50 best simulations. While the optimal configuration yields a selected charge comparable to that of the 50 best configurations, its spectrum exhibits a more pronounced QMEP centered near $46\,\si{\MeV}$, compared with a broader peak centered near $42\,\si{\MeV}$ in the averaged spectrum. The higher concentration of charge at higher energies is consistent with the greater laser-to-beam energy conversion efficiency achieved by the optimal configuration.

\begin{figure*}[!t]
    \centering
    \includegraphics[scale=1]{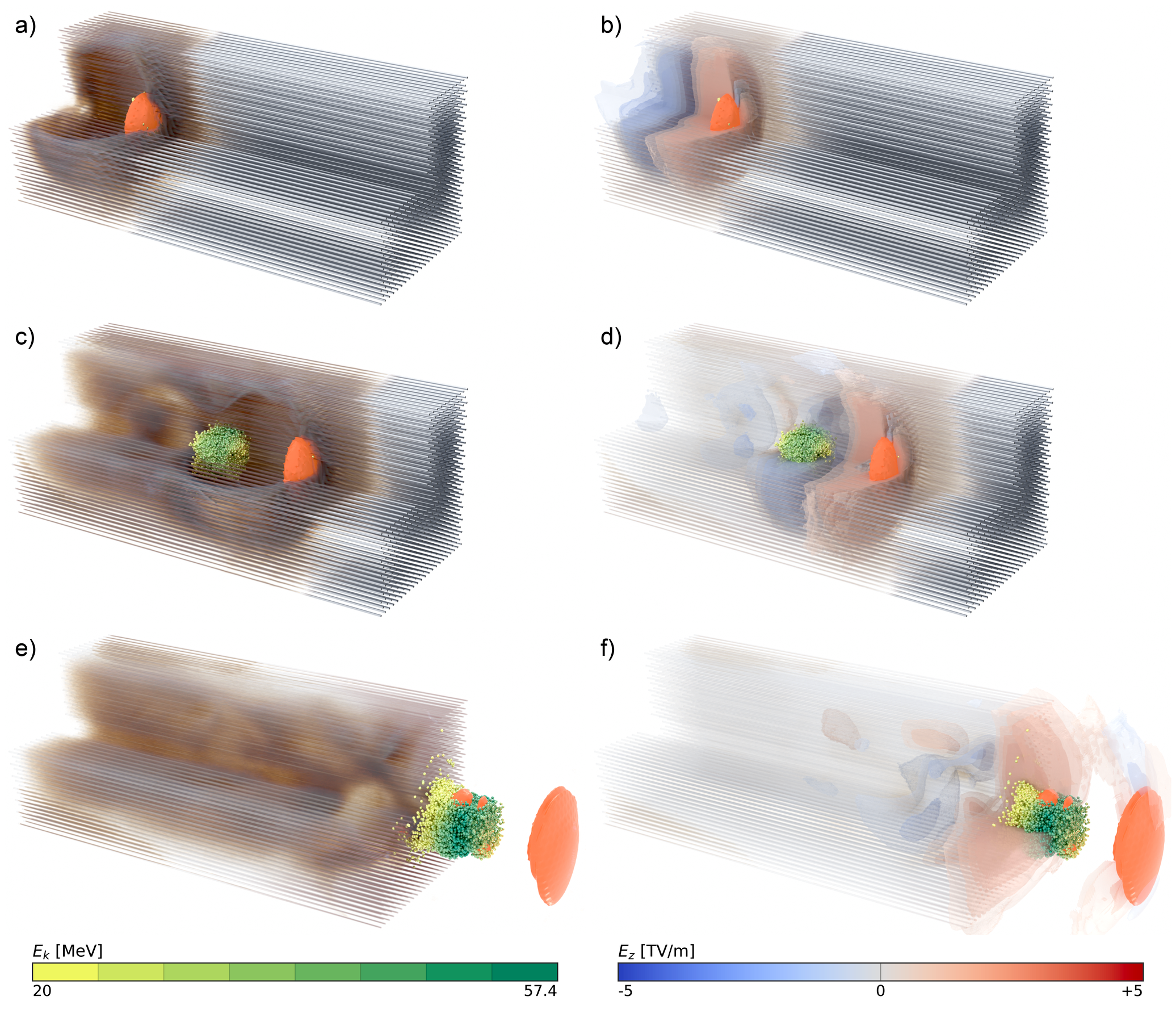}
    \caption{Three-dimensional evolution of laser wakefield acceleration in the optimal LA-CNT target of the first optimization campaign ($O_1$), at early bubble formation (a,b), during acceleration inside the target (c,d), and at bunch extraction near the target exit (e,f). CNT bundles are drawn as longitudinal rods, sharply defined while neutral and much fainter once ionized, so that the fading of the array marks the ionized region. The translucent haze is the plasma density (electrons and carbon ions, not distinguished by colour), dominated by the electrons, which outline the walls of the wakefield cavity. The orange surface is the laser pulse. Electrons with instantaneous kinetic energies $E_k>20\,\si{\MeV}$ are displayed as spheres coloured according to their kinetic energy (yellow--green scale, in \si{\MeV}), independently selected at each snapshot without backtracking from the final bunch. Panels (b,d,f) correspond to the same snapshots as (a,c,e), respectively, with the plasma-density visualization replaced by isosurfaces of the longitudinal electric field $E_z$ (blue--white--red scale, limited to $\pm 5\,\si{\tera\volt/\meter}$ for improved visualization).}
    \label{fig:results3d}
\end{figure*}

The underlying physics of the acceleration process can be visualized in Fig.~\ref{fig:results3d}, which presents three-dimensional snapshots of the laser--target interaction for the optimal configuration at three stages of propagation: shortly after the laser enters the target (a,b), near the target center (c,d), and during bunch extraction (e,f). As the laser pulse propagates, the CNT bundles are ionized and the released electrons are expelled away from the propagation axis, forming a wakefield bubble that travels through the target~\cite{Bontoiu2025,Nunes2026,Bonatto2026}. This process resembles the bubble regime observed in conventional LWFA in gaseous plasmas and in hollow-channel plasma acceleration~\cite{Gessner2016}, as discussed in previous studies~\cite{Bonatto2023,Bontoiu2025}.

The left column illustrates the evolution of the laser pulse (orange isosurface), the CNT bundles (longitudinal rods), and the electron population with instantaneous kinetic energies above $20\,\si{\MeV}$ (yellow--green scale). At the earliest stage, Fig.~\ref{fig:results3d}(a), few electrons have reached this energy threshold. The progressive fading of the CNT bundles indicates the advance of the ionization front, while the translucent plasma-density distribution reveals the formation of the wakefield cavity, particularly evident in panel (c). Although electrons and carbon ions are not distinguished by colour, the ions remain approximately at their original positions over the femtosecond timescale of the interaction, undergoing only limited expansion due to Coulomb repulsion following electron removal~\cite{Bontoiu2023}. Consequently, the evolving cavity and its surrounding plasma sheath are primarily associated with the redistribution of electrons. The right column shows the corresponding longitudinal electric field, $E_z$, represented by blue--white--red isosurfaces. The characteristic field structure of the bubble regime is observed, with a positive, decelerating field in the front region of the cavity and a negative, accelerating field in its rear region, where electrons are trapped and accelerated. The accelerating field reaches the TV/m scale, and the electron bunch remains within the accelerating phase of the wake as it propagates through the target, as illustrated in panels (b) and (d). Near the target exit, the wakefield structure extends beyond the nanostructured medium, and the bunch begins to encounter regions of positive $E_z$, as shown in panel (f).

\begin{figure*}[!t]
    \centering
    \includegraphics[scale=1]{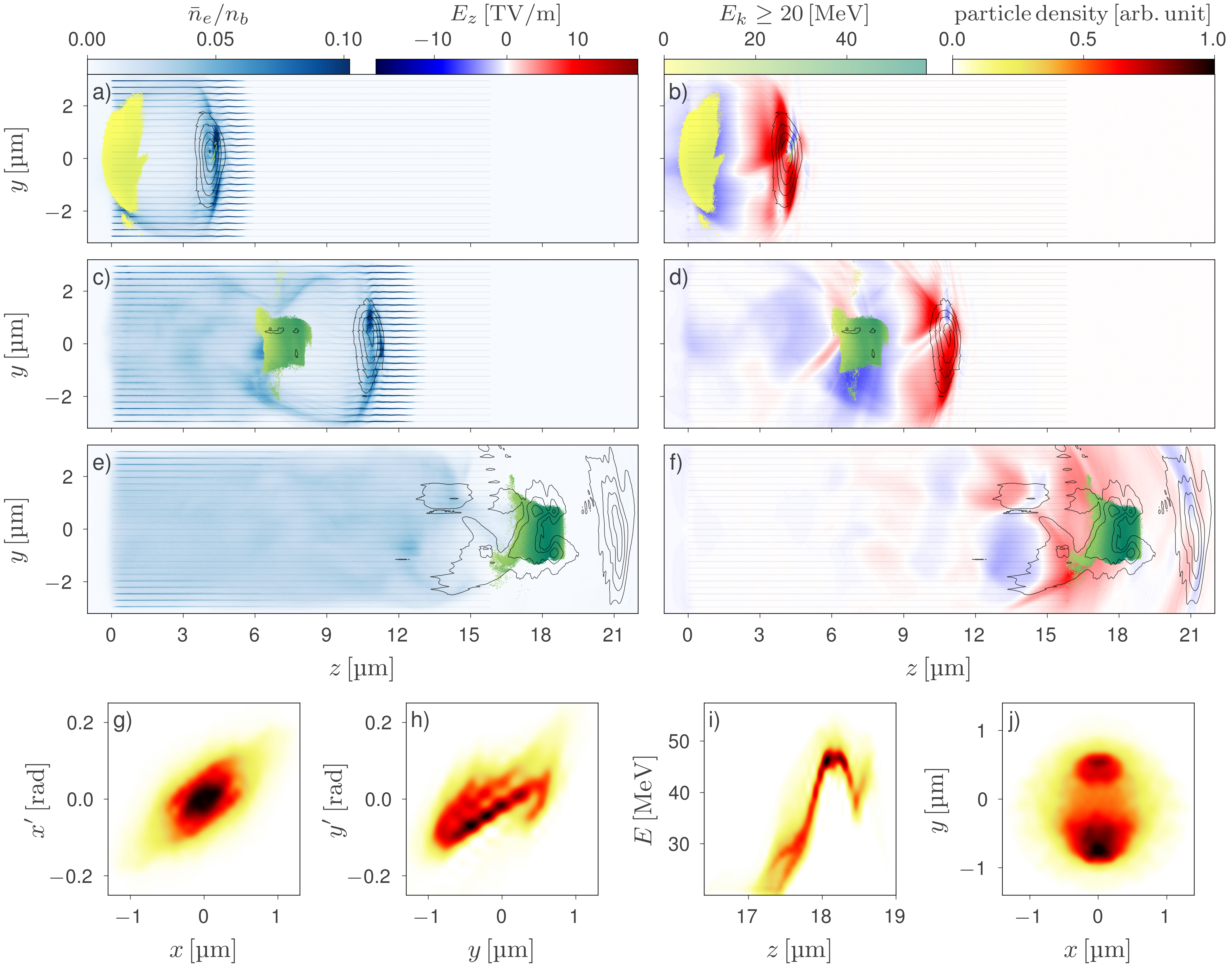}
    \caption{$O_1$ optimal case. Electron density averaged along the $x$ direction (blue colormap, saturated to enhance the visibility of the wakefield-bubble contour), longitudinal electric field (blue–white–red colormap), and accelerated electron macroparticle energy (yellow–green colormap) for the optimal input configuration. Panels (a)–(f) correspond to the same three stages of propagation as Fig.~\ref{fig:results3d}. In panels (a)–(f), the yellow–green markers show the macroparticles selected at the final time step and backtracked to the displayed times; gray lines indicate a subset of the CNT bundles, and black contours the laser intensity. Panels (g)–(i) show the transverse and longitudinal phase spaces at extraction, and panel (j) shows the spatial distribution of the accelerated electron beam. In panels (g--j) the white–yellow–red–black colormap indicates the normalized particle density.}
    \label{fig:results}
\end{figure*}

A complementary, quantitative view of the same process is presented in Fig.~\ref{fig:results}. Panels (a), (c), and (e) show the electron density averaged along the $x$ direction (blue scale), saturated to emphasize the bubble contour. Panels (b), (d), and (f) show the longitudinal electric field (blue--white--red scale) at the same three stages of propagation. In all panels, light-gray lines indicate a subset of the CNT bundles, and black contours represent the laser intensity. A key difference from Fig.~\ref{fig:results3d} is that the accelerated electron bunch is defined by the particles extracted from the target with longitudinal momenta corresponding to kinetic energies above $20\,\si{\MeV}$, as described in Sec.~\ref{sec:pic}. These particles (yellow--green scale) are identified at the end of the simulation and backtracked throughout the interaction. Consequently, electrons that eventually form the accelerated bunch are already visible at earlier stages, even before their kinetic energies exceed the aforementioned threshold.

Figures~\ref{fig:results}(a,b) show the early stage of acceleration, with the laser pulse located at approximately $z\simeq5\,\si{\um}$ and longitudinal electric fields exceeding $10\,\si{\tera\volt/\meter}$ within the wakefield bubble. Even at this early stage, electrons that will contribute to the final accelerated bunch have already been self-injected at the rear of the bubble, potentially allowing them to experience acceleration over a substantial fraction of the target length. As the bubble propagates towards the center of the target, Figs.~\ref{fig:results}(c,d) show the continued acceleration of trapped electrons, with those near the centre of the bunch reaching energies of up to $45\,\si{\MeV}$, exceeding those at its head and tail. Near the target exit, Figs.~\ref{fig:results}(e,f) show the accelerated electron bunch emerging from the target.

Figures~\ref{fig:results}(g)--(j) characterize the extracted electron bunch. The transverse phase spaces ($x$--$x'$ and $y$--$y'$, where $x'=u_x/u_z$ and $y'=u_y/u_z$) in panels (g) and (h) reveal similar RMS divergences along both axes ($\sim71\,\si{\milli\radian}$ in $x$ and $\sim70\,\si{\milli\radian}$ in $y$). However, the distribution is more symmetric along the $x$ direction, whereas a slight asymmetry and offset are observed along the $y$ direction, parallel to the laser polarization, consistent with previously reported polarization effects in LA-CNT targets~\cite{Nunes2026}. The longitudinal phase space ($z$--$E_k$) in panel (i) shows that the bunch extends over approximately $1.5\,\si{\um}$, with a charge of $673\,\si{\pico\coulomb}$, a mean energy of $36.6\,\si{\MeV}$, an RMS energy spread of $8.6\,\si{\MeV}$, and a maximum energy of $56.7\,\si{\MeV}$. Finally, Fig.~\ref{fig:results}(j) shows the transverse spatial distribution of the bunch, with a larger spread along the $y$ direction and a higher particle density in its lower region. Most of the bunch charge is concentrated within a transverse area of approximately $4\,\si{\um\squared}$ immediately after exiting the target.

\begin{figure}[!t]
    \centering
    \includegraphics[scale=1]{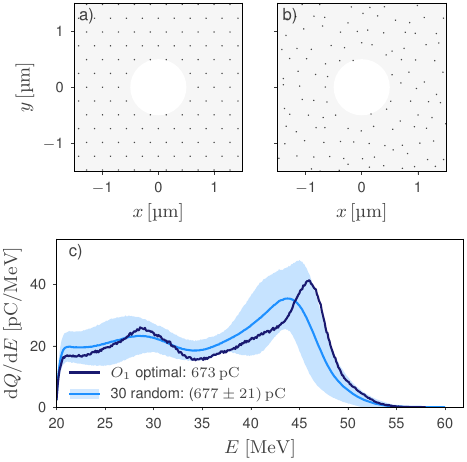}
    \caption{CNT targets showing the central region of (a) the $O_1$ optimal configuration and (b) an example of a perturbed configuration obtained by introducing Gaussian noise with $\sigma = 4r_b$ to the CNT bundle positions. In both panels, bundles whose centres lie within the white central circle of radius $r_\mathrm{in}$ were removed. (c) Average electron energy spectrum from 30 simulations (light blue), with $\pm 2s$ shading, compared with the spectrum of the optimal configuration (dark blue).}
    \label{fig:results_random}
\end{figure}

To assess the robustness of the system to imperfections in target fabrication and its sensitivity to deviations from the optimal target configuration, shown in Fig.~\ref{fig:results_random}(a), Gaussian random displacements with standard deviation $\sigma$ were added to the positions of the CNT bundles. In a preliminary set of simulations, $\sigma$ was progressively increased from $r_b$ to $4r_b$, in steps of $r_b$. As similar results were obtained for all perturbation levels, only the largest value, $\sigma = 4r_b$, is presented here. Thirty simulations were then performed with this value of $\sigma$, each generated with a different random seed. The resulting bundle distributions are highly irregular, with no discernible pattern, as illustrated in Fig.~\ref{fig:results_random}(b). Since the bundles lying inside the aperture are removed only after the random displacements are applied, the number of removed bundles differed between realizations, so that the total number of CNT bundles ranged from 525 to 529 across the simulations. This corresponds to a relative variation of $(0.2 \pm 0.1)\%$ in the effective density compared with the optimal configuration.

Figure~\ref{fig:results_random}(c) shows the average of the 30 energy spectra obtained with the perturbed targets and their $\pm s$ band, as well as the spectrum of the optimal target (dark blue line). The total beam charge and the conversion efficiency are essentially preserved, averaging $(677 \pm 21)$~\si{\pico\coulomb} and $(24.2 \pm 0.7)\%$, respectively, compared with 673~\si{\pico\coulomb} and 24.6\% for the unperturbed target. The main effect of the disorder is on the high-energy part of the spectrum, whose peak shifts from $46$~\si{\MeV} in the optimal case to $43.8$~\si{\MeV} in the averaged spectrum, a reduction of about $5\%$, although the optimal spectrum still lies mostly within the shaded band.

This contrast is consistent with the effective-density approach. The charge and efficiency are expected to be set mainly by $n_{\mathrm{eff}}$, which is nearly unchanged. The spectral peak, in contrast, also depends on the local arrangement of the bundles near the axis, where the laser intensity is highest and the wakefield is driven. In this region, the random displacements and the removal or addition of bundles at the aperture modify the structure sampled by the pulse, even though the overall density is preserved.

A further aspect of the optimization results concerns the bounds of the search ranges listed in Table~\ref{tab:input}. Figures~\ref{fig:results_bayesian}(b–d) show that the algorithm frequently suggested points near these bounds. As discussed above, increasing $L$ further would lead to GPU memory limitations, whereas reducing $r_b$ would compromise the reliability of the results because of insufficient numerical resolution. The aperture radius is a different case. Figure~\ref{fig:results_bayesian}(d) shows that the algorithm repeatedly explored the minimum allowed value, $r_{\mathrm{in}} = 0.5~\si{\um}$, which consistently yielded high values of $\eta$. This behaviour suggests that the optimum may lie below the bound. Unlike the bounds on $L$ and $r_b$, this one can be lowered to zero without any computational difficulty.

The limiting value $r_{\mathrm{in}} = 0$ corresponds to a target with no central aperture and, consequently, with no need to remove CNT bundles after fabrication. Such an aperture-free target would also eliminate the experimental challenges of fabricating the aperture and aligning the laser pulse within it, which makes it particularly attractive for a proof-of-concept experiment. A second optimization ($O_2$) was therefore performed, identical to the first one ($O_1$) except for the range of $r_{\mathrm{in}}$, which was extended to $0 \leq r_{\mathrm{in}} \leq 2.5~\si{\um}$. Its aim was to verify whether a better optimum exists below the previous bound on $r_{\mathrm{in}}$ and, in particular, whether aperture-free targets can match the efficiency of $O_1$ while retaining the features that make the $O_1$ spectra attractive, notably a pronounced QMEP at high energy. Although such features are not explicitly rewarded by the objective function, they are as relevant to the practical value of a target as the efficiency itself.

\begin{figure}[!t]
    \centering
     \includegraphics[scale=1]{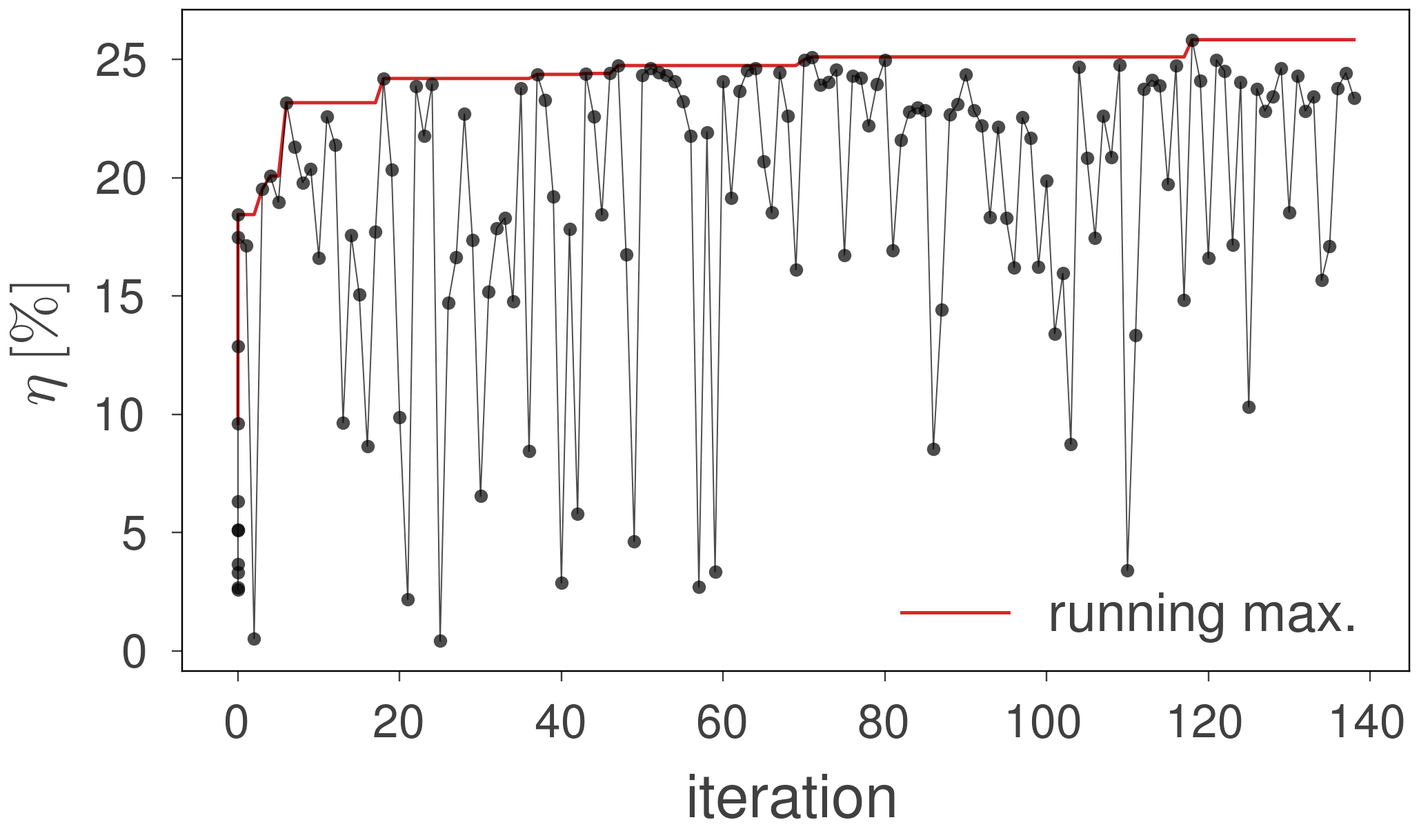}
    \caption{Evolution of the objective function $\eta$ over the course of $O_2$. Black markers show $\eta$ at each iteration, and the red line indicates the best value obtained up to that point.}
    \label{fig:results_iterationo2}
\end{figure}

Figure~\ref{fig:results_iterationo2} shows the evolution of $\eta$ over the 138 iterations of $O_2$, with 10 initial random evaluations grouped at iteration 0. The pattern closely follows that of $O_1$. Within about 20 iterations the algorithm identifies a region of high efficiency, and it then alternates between exploring other parts of the parameter space and exploiting the vicinity of this optimum. The highest value, $\eta = 25.8\%$, is reached at iteration 118, with a relative increase of about 5\% over the 24.6\% obtained for the optimal case in $O_1$.

\begin{figure*}[!t]
    \centering
    \includegraphics[scale=1]{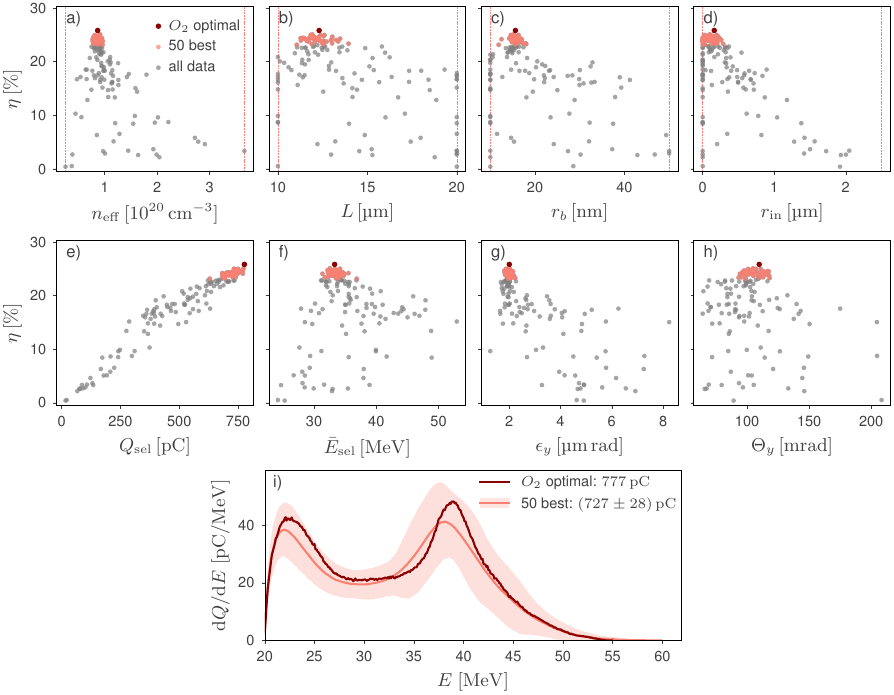}
    \caption{Scatter plots of the $O_2$ Bayesian optimization. The dark red point is the simulation that resulted in the highest $\eta$ value, light red points indicate the 50 best subset and gray points are the remaining simulations. Input parameters: $\eta$ as a function of (a) the effective target density, (b) the target length, (c) the CNT bundle radius, and (d) the central aperture radius. Output beam parameters---considering only particles with energies above 20~\si{\MeV}: $\eta$ as a function of (e) beam charge, (f) average energy, (g) transverse emittance in the $y$ direction, and (h) beam divergence in the $y$ direction. (i) Electron energy spectra corresponding to the optimal simulation (dark red) and to the average of the 50 best (light red). Shading indicates the mean $\pm 2s$.}
    \label{fig:results_bayesiano2}
\end{figure*}

Figures~\ref{fig:results_bayesiano2}(a)–(d) show, for $O_2$, scatter plots of $\eta$ as a function of the input parameters, and panels (e)–(h) the corresponding plots for the output parameters. Figure~\ref{fig:results_bayesiano2}(a) shows that the 50 best configurations cluster around $n_\mathrm{eff} = (8.6 \pm 0.5)\times10^{19}\,\si{\per\cm\cubed}$, about 21\% above the corresponding cluster in $O_1$. In contrast, Fig.~\ref{fig:results_bayesiano2}(b) shows that they cluster around $L = (12.2 \pm 0.6)\,\si{\um}$, below the 14--20~\si{\um} range of high efficiency found in $O_1$. The bundle radius increases slightly, to $r_b = (15 \pm 1)\,\si{\nm}$. Finally, Fig.~\ref{fig:results_bayesiano2}(d) shows that the aperture radius clusters at $r_{\mathrm{in}} = (0.11 \pm 0.07)\,\si{\um}$, well below the lower bound of $O_1$, confirming that the optimum lies in the region excluded from the first campaign.

Figures~\ref{fig:results_bayesiano2}(e) and (f) show that the 50 best configurations of $O_2$ yield beams with charges of $(727 \pm 28)\,\si{\pico\coulomb}$ and mean energies of $(33.4 \pm 1.0)\,\si{\MeV}$. Compared with $O_1$, the charge is about 7\% higher and the mean energy about 5\% lower, with a net relative increase of 1.3\% in the average efficiency. This modest gain is accompanied by reduced beam quality. Because $O_2$ favours smaller apertures and higher effective densities, more CNT bundles are located near the target centre, where the laser intensity is highest. This hinders laser propagation through the target and increases both the emittance and the divergence of the beam, as shown in Figs.~\ref{fig:results_bayesiano2}(g,h). Their averages increase by 11\% and 48\%, respectively, relative to the 50 best configurations of $O_1$. The spectra in Fig.~\ref{fig:results_bayesiano2}(i) show that the optimal configuration produces a beam with a higher charge (777~\si{\pico\coulomb}) and a more pronounced QMEP at a slightly higher energy than in the averaged spectrum. In both spectra, however, the QMEP lies near 38~\si{\MeV}, below the 42~\si{\MeV} of the averaged $O_1$ spectrum and the 46~\si{\MeV} of the $O_1$ optimum. The higher charge obtained with $O_2$ is thus accompanied by a lower QMEP energy, reinforcing the trade-off discussed below.

\begin{figure*}[!t]
    \centering
    \includegraphics[scale=1]{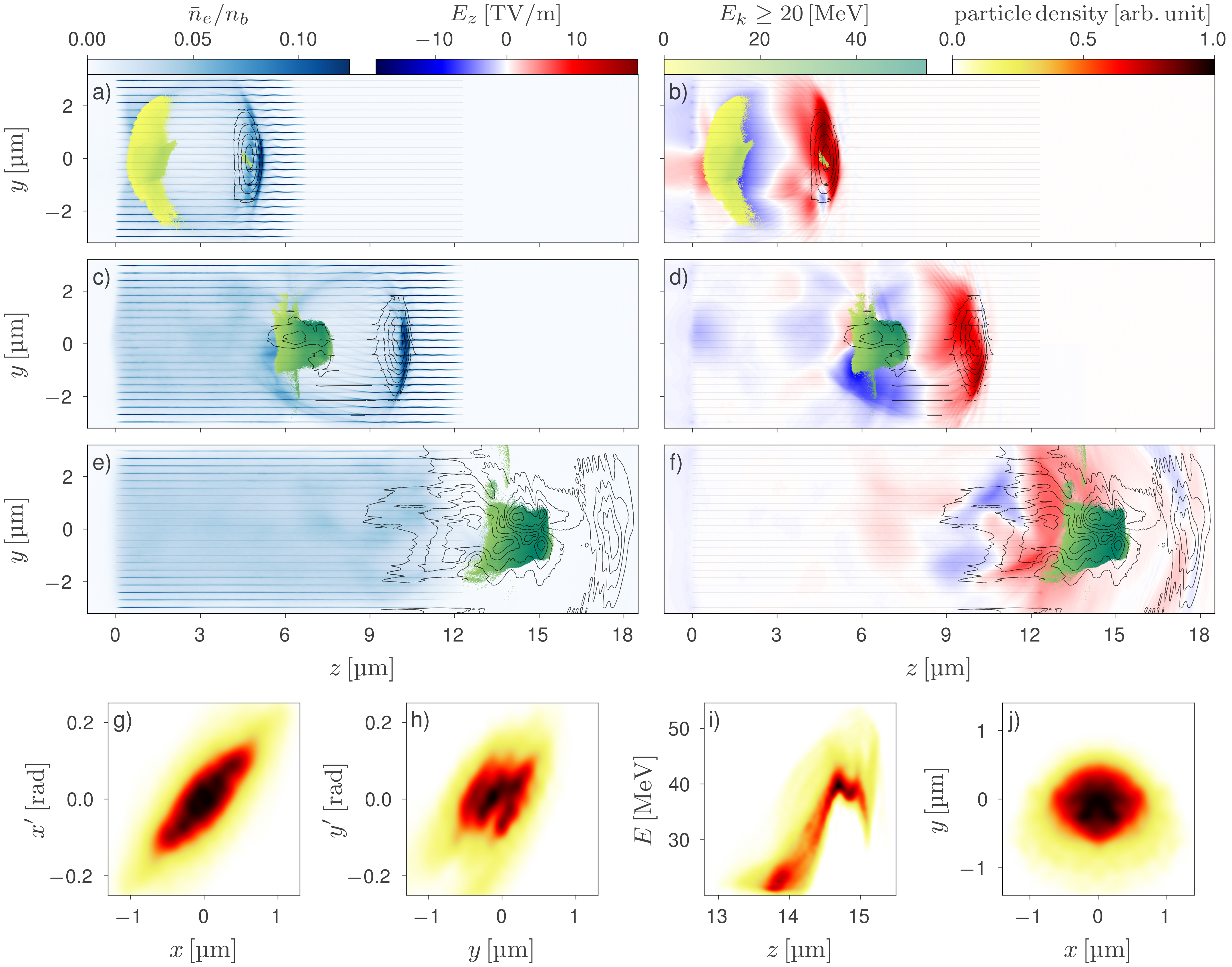}
    \caption{Electron density averaged along the $x$ direction (blue colormap), longitudinal electric field (blue–white–red colormap), and accelerated electron macroparticle energy (yellow–green colormap) for the optimal input configuration. The electron-density colormap is saturated to enhance the visibility of the wakefield-bubble contour. Panels (a)–(f) correspond to different stages of wakefield bubble propagation. Panels (g)–(i) show the transverse and longitudinal phase spaces at extraction, and panel (j) shows the spatial distribution of the accelerated electron beam.  In panels (g--j) the white–yellow–red–black colormap indicates the normalized particle density.}
    \label{fig:resultso2}
\end{figure*}

Figure~\ref{fig:resultso2}(a–f) presents three snapshots of the optimal $O_2$ simulation. Overall, the dynamics are very similar to those observed in Fig.~\ref{fig:results}, with a well-defined wakefield bubble propagating through the target, and a higher-energy electron population concentrated around $40~\si{\MeV}$. However, some differences can be identified. The first is that the laser intensity contours in Fig.~\ref{fig:resultso2}(e,f) are noticeably more distorted than those in Fig.~\ref{fig:results}(e,f). The smaller central aperture places more CNT bundles within the high-intensity region of the pulse, enhancing the laser–target interaction and disrupting pulse propagation, as discussed above. This increases the beam charge but lowers the QMEP energy and degrades the beam quality. As shown in Figs.~\ref{fig:resultso2}(g,h), the beam divergence is more than $40\%$ higher in both transverse directions than in the optimal $O_1$ case shown in Figs.~\ref{fig:results}(g,h). The longitudinal phase space in Fig.~\ref{fig:resultso2}(i) also reveals a broader energy distribution, with more low-energy electrons. Despite these changes, the accelerated beam is more spatially centred, as shown in Fig.~\ref{fig:resultso2}(j).

A noteworthy feature of the $O_2$ results is that some of the 50 best configurations have near-zero values of $r_{\mathrm{in}}$, as shown in Fig.~\ref{fig:results_bayesiano2}(d). Such values can yield aperture-free targets, provided that no bundle centre lies within the prescribed aperture, so that none is removed. Given the practical advantages of aperture-free targets discussed above, a separate analysis of their performance is performed. Selecting only configurations in which no bundles were removed from the 50 best results of $O_2$ yields a subset of 23 cases, hereafter denoted by $O_2^*$. Since the optimal $O_2$ configuration, whose dynamics are shown in Fig.~\ref{fig:resultso2}, required the removal of one bundle, it is excluded from this subset.

\begin{figure}[!t]
    \centering
    \includegraphics[scale=1]{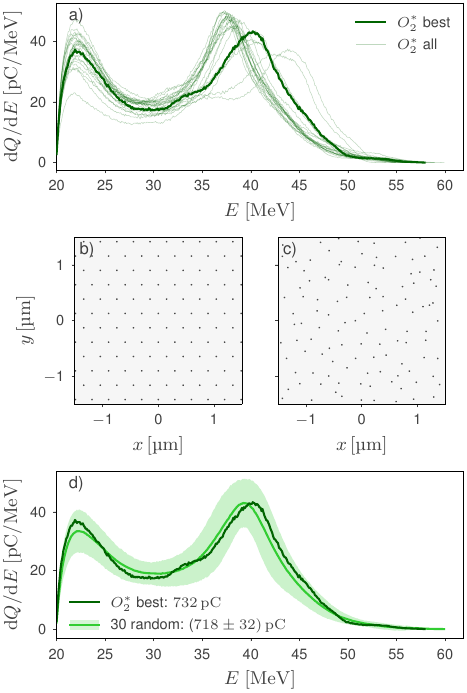}
    \caption{(a) Electron energy spectra for the 23 aperture-free configurations in $O_2^*$, with the highest-efficiency (best) case, $\eta \simeq 25\%$, highlighted by a thicker dark-green line. Central region of the transverse CNT bundle arrangements for (b) this configuration and (c) one of 30 perturbed targets obtained by adding Gaussian displacements with $\sigma = 4r_b$ to the bundle positions. (d) Average electron energy spectrum of the 30 perturbed simulations (light green), with $\pm 2s$ shading, compared with the spectrum of the unperturbed configuration (dark green).}
    \label{fig:results_random2}
\end{figure}

Figure~\ref{fig:results_random2}(a) shows the electron energy spectra of the 23 configurations in $O_2^*$. The spectra exhibit a similar two-peak structure, with a low-energy peak near 22~\si{\MeV} and a higher-energy QMEP that lies approximately between 36 and 40~\si{\MeV} in most configurations, with variations in its position and amplitude. The highest-efficiency configuration ($O_2^*$ best), highlighted in dark green, has its QMEP near 40~\si{\MeV}. To assess the robustness of this case against positional disorder, 30 simulations were performed with Gaussian random displacements of standard deviation $\sigma = 4r_b$ applied to the CNT bundle positions. Figures~\ref{fig:results_random2}(b,c) show the central regions of the unperturbed target and one of the 30 perturbed targets, respectively. Figure~\ref{fig:results_random2}(d) compares the spectrum of the best-case configuration with the average spectrum of its 30 perturbed realizations and the corresponding shaded band. The unperturbed spectrum lies largely within this band, with only minor spectral changes, including a shift of approximately 1~\si{\MeV} in the QMEP energy. These results indicate that the beam properties are robust against fluctuations in the CNT positions, even in the absence of a central aperture.

\begin{figure}[!t]
    \centering
    \includegraphics[scale=1]{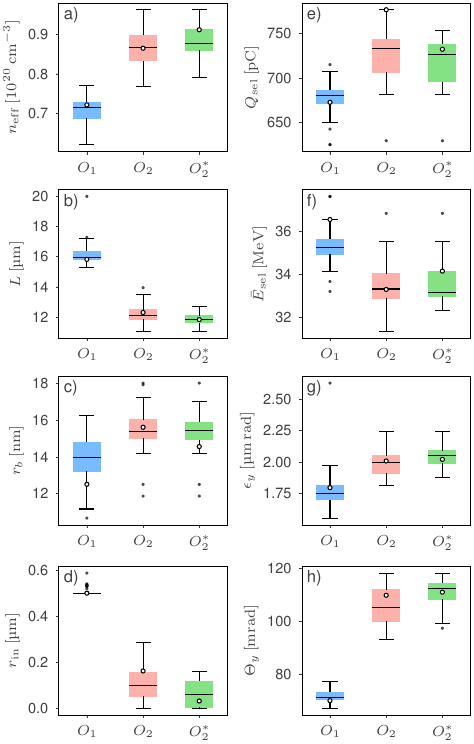}
    \caption{Boxplots of the input parameters: (a) effective density, (b) target length, (c) bundle radius and (d) aperture radius; and the output parameters: (e) selected charge, (f) mean energy, (g) normalized emittance in the $y$ direction and (h) divergence in the $y$ direction. The distributions correspond to the 50 best configurations of $O_1$, the 50 best configurations of $O_2$ and the 23 configurations in $O_2^*$. In each boxplot, a white marker indicates the value corresponding to the highest-$\eta$ simulation; black markers indicate outliers.}
    \label{fig:boxplots}
\end{figure}

Figure~\ref{fig:boxplots} compares the target parameters and resulting beam properties across the 50 best configurations of $O_1$ and $O_2$ and the aperture-free subset $O_2^*$. The input distributions reveal distinct preferred geometries. Compared with $O_1$, both $O_2$ groups favour higher effective densities and shorter targets [panels (a,b)], together with slightly larger bundle radii and much smaller nominal aperture radii [panels (c,d)]. The distributions of $O_2$ and $O_2^*$ largely overlap, although the aperture-free subset tends towards slightly shorter targets and smaller aperture radii. These geometric differences are accompanied by a clear trade-off in beam properties. Both $O_2$ groups yield higher charges but lower mean energies than $O_1$ [panels (e,f)]. The normalized emittance in the $y$ direction increases moderately, whereas the divergence rises substantially, from approximately 70~\si{\milli\radian} in $O_1$ to 105--110~\si{\milli\radian} in the two $O_2$ groups [panels (g,h)]. These trends are consistent with the interaction dynamics discussed above: the larger aperture and lower effective density of $O_1$ favour sustained laser propagation and acceleration, whereas the greater charge obtained in $O_2$ is accompanied by lower electron energies and increased transverse spreading. The similar output distributions of $O_2$ and $O_2^*$ show that aperture-free targets retain the characteristic beam properties of the broader $O_2$ group while offering the practical advantages discussed earlier.

\begin{figure}[!t]
    \centering
    \includegraphics[scale=1]{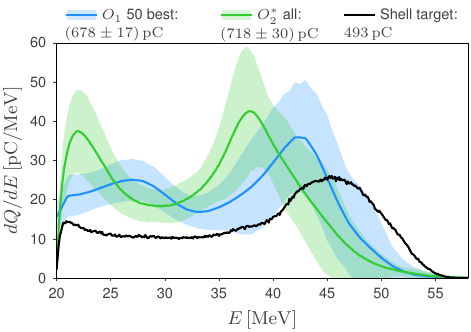}
    \caption{Average electron energy spectra for the $O_1$, in blue, $O_2^*$, in green, and the shell target\cite{Bonatto2026} in black. Shading indicates the mean $\pm 2s$.}
    \label{fig:results_cristian}
\end{figure}

The performance of the optimized staggered LA-CNT targets is further assessed against that of the concentric-shell target introduced in Ref.~\cite{Bontoiu2025}. Figure~\ref{fig:results_cristian} presents the average electron energy spectra of the 50 best configurations of $O_1$ and the complete subset of $O_2^*$, with their respective shaded bands, together with the spectrum of the concentric-shell target. For this comparison, the latter was re-simulated with the numerical parameters of Table~\ref{tab:PIC} and the post-processing procedure of the present work, as described in Ref.~\cite{Bonatto2026}. The optimized configurations yield laser-to-beam energy conversion efficiencies of $\eta \approx 24\%$, compared with $\eta = 19\%$ for the concentric-shell target (black curve), a representative case selected from a preliminary parameter scan rather than an optimized design~\cite{Bonatto2026}. The comparison thus shows that the simpler geometry matches or exceeds the efficiency of the shell design, although part of the difference may reflect the broader search of the Bayesian optimization. The higher efficiency is primarily associated with the higher accelerated charge, since the average charges in $O_1$ and $O_2^*$ exceed that of the concentric-shell target by approximately 185~\si{\pico\coulomb} and 225~\si{\pico\coulomb}, respectively.

The beam properties of the two optimized groups and the concentric-shell reference target are summarized in Table~\ref{tab:comparison}. Both optimized groups achieve similar conversion efficiencies, exceeding that of the reference, and comparable normalized emittances, below the reference values. Their averaged spectra exhibit two peaks, with the higher-energy QMEPs centred at lower energies than the single broad peak of the reference spectrum. $O_1$ also yields lower divergences than the reference, whereas $O_2^*$ combines the highest estimated accelerating gradient with approximately 40\% larger divergences. Despite the significant increase in beam divergence, the simpler fabrication and less demanding alignment and repositioning of aperture-free targets may make $O_2^*$ an attractive alternative for proof-of-concept experiments and applications that can tolerate reduced beam quality.

\begin{table}[!t]
\caption{\label{tab:comparison}Beam properties of the 50 best $O_1$ configurations, the $O_2^*$ subset and the concentric-shell reference target. For the optimized groups, peak energies refer to the averaged spectra, whereas the remaining entries are means $\pm$ standard deviations across the respective sets. Accelerating gradients are estimated from the highest-energy spectral peak. Here, $\epsilon$ and $\Theta$ denote the RMS normalized emittance and RMS divergence, respectively.}
\centering
\begin{tabular*}{0.94\columnwidth}{@{\extracolsep{\fill}}ccccc@{}}
\hline
\multirow{2}{*}{\makecell[c]{Quantity}} & \multirow{2}{*}{\makecell[c]{Unit}} & \multirow{2}{*}{\makecell[c]{$O_1$}} & \multirow{2}{*}{\makecell[c]{$O_2^*$}} & \multirow{2}{*}{\makecell[c]{Shell\\target}}\\
&&&&\\
\hline\\[-2ex]
$\eta$ & \% & $23.9 \pm 0.3$ & $24.1 \pm 0.5$ & 19.0\\[1.0ex]
$Q_{\mathrm{sel}}$ & \si{\pico\coulomb} & $678 \pm 17$ & $718 \pm 30$ & 493\\[1.0ex]
Main peak & \si{\MeV} & 42.2 & 37.8 & 45.3\\[1.0ex]
Second peak & \si{\MeV} & $\sim 27$ & $\sim 22$ & ---\\[1.0ex]
Gradient & \si{\tera\eV/\meter} & $2.6 \pm 0.2$ & $3.3 \pm 0.2$ & 3.0\\[1.0ex]
$\epsilon_x$ & \si{\micro\meter\radian} & $1.8 \pm 0.1$ & $1.8 \pm 0.1$ & 2.6\\[1.0ex]
$\epsilon_y$ & \si{\micro\meter\radian} & $1.8 \pm 0.2$ & $2.0 \pm 0.1$ & 2.5\\[1.0ex]
$\Theta_x$ & \si{\milli\radian} & $73 \pm 3$ & $106 \pm 5$ & 79\\[1.0ex]
$\Theta_y$ & \si{\milli\radian} & $72 \pm 2$ & $111 \pm 6$ & 78\\
\hline
\end{tabular*}
\end{table}

\section{Conclusion and Discussion}\label{conclusion}
A simplified LA-CNT target architecture has been proposed and optimized for LWFA driven by a 100~\si{\milli\joule}, few-cycle, 800~\si{\nm} laser pulse, as an alternative to the concentric-shell targets of Refs.~\cite{Bontoiu2025,Nunes2026,Bonatto2026}. Its principal advantage is the removal of the requirement to place CNT bundles on prescribed concentric shells, which relaxes the geometric constraints on target fabrication while retaining efficient bubble-regime acceleration. The CNT bundles are aligned along the laser propagation direction and arranged in staggered rows forming a triangular lattice within a square transverse cross section, with an optional central aperture. Bayesian optimization coupled to fully three-dimensional WarpX PIC simulations was used to vary the effective density, target length, bundle radius and aperture radius while keeping all laser parameters fixed. The objective was to maximize the laser-to-beam energy conversion efficiency for extracted electrons whose forward longitudinal momenta correspond to kinetic energies above 20~\si{\MeV}, with charge and mean energy entering the objective through their product.

In the first campaign ($O_1$), with the aperture radius constrained to $r_{\mathrm{in}}\geq0.5~\si{\micro\meter}$, the best configuration reached an efficiency of 24.6\%, delivering a charge of 673~\si{\pico\coulomb}, a mean energy of 36.6~\si{\MeV} and a QMEP near 46~\si{\MeV} in a well-defined wakefield bubble. The 50 best configurations favour lower effective densities, longer targets and apertures near the lower bound. They yield a mean efficiency of $(23.9\pm0.3)\%$, a mean charge of $(678\pm17)~\si{\pico\coulomb}$, a QMEP near 42~\si{\MeV} in their averaged spectrum and RMS divergences of approximately 72--73~\si{\milli\radian}.

Since $O_1$ repeatedly favoured the smallest admissible aperture, the second campaign ($O_2$) allowed $r_{\mathrm{in}}$ to reach zero. The optimization favoured denser and shorter targets with much smaller apertures, and the maximum efficiency increased to 25.8\%, although that configuration still required the removal of one bundle. The 50 best $O_2$ configurations yield a mean charge of $(727\pm28)~\si{\pico\coulomb}$ and a mean energy of $(33.4\pm1.0)~\si{\MeV}$. Among them, the 23 aperture-free configurations requiring no bundle removal form the subset $O_2^*$. This subset retains a mean efficiency of $(24.1\pm0.5)\%$ and a mean charge of $(718\pm30)~\si{\pico\coulomb}$, with a higher-energy QMEP at 37.8~\si{\MeV} in its averaged spectrum. Its emittances are comparable to those of $O_1$, while its divergences of 106--111~\si{\milli\radian} are approximately 50\% larger.

These differences are consistent with the laser--target interaction dynamics observed in the simulations. The larger aperture and lower effective density of $O_1$ favour sustained pulse propagation and acceleration over longer distances. With smaller or absent apertures, more bundles occupy the high-intensity region near the axis. The increased interaction can enhance self-injection and charge, while pulse deformation limits the useful acceleration length. The preference for shorter targets is consistent with extracting the bunch before further wakefield deterioration. The two regimes can attain similar objective values despite their different QMEP energies and divergences, which are not explicitly optimized.

The reduction in geometric complexity comes at no cost in efficiency. Under the same numerical resolution and particle-selection criterion, the concentric-shell target, a representative case selected from a preliminary parameter scan rather than an optimized design~\cite{Bonatto2026}, reaches 19\%, whereas the optimized LA-CNT groups reach approximately 24\%, mainly owing to higher accelerated charges, which exceed the reference by approximately 185~\si{\pico\coulomb} in $O_1$ and 225~\si{\pico\coulomb} in $O_2^*$. The shell spectrum has a single broad peak at 45.3~\si{\MeV}, above the higher-energy QMEPs of the averaged optimized spectra. The estimated accelerating gradients are $(2.6\pm0.2)$, $(3.3\pm0.2)$ and 3.0~\si{\tera\eV/\meter} for $O_1$, $O_2^*$ and the shell target, respectively. Both optimized groups have lower normalized emittances than the reference, and $O_1$ also has a lower divergence, whereas that of $O_2^*$ is substantially higher.

Sensitivity to positional disorder was assessed using 30 realizations of Gaussian bundle displacements with $\sigma=4r_b$ for each of the best configurations of $O_1$ and $O_2^*$, with nearly unchanged effective density. For the tested $O_1$ configuration, the mean efficiency under perturbation is $(24.2\pm0.7)\%$, compared with 24.6\% before perturbation, and the charge is largely retained. The QMEP of the averaged spectrum shifts from approximately 46 to 43.8~\si{\MeV}. The corresponding shift for the tested $O_2^*$ configuration is approximately 1~\si{\MeV}, also with modest changes in charge and efficiency. These results support tolerance to the tested positional disorder in the two configurations while showing that high-energy spectral features retain a dependence on the local bundle arrangement.

These findings extend earlier effective-density approaches involving concentric plasma tubes, graphene layers and CNT bundles on concentric shells to staggered LA-CNT arrays. Within the investigated model and parameter ranges, they support effective density as a key design parameter for efficient bubble-regime acceleration, while showing that aperture geometry and local bundle arrangement remain important for spectral and transverse beam properties. This design freedom is central to the practical value of the work: high conversion efficiencies are obtained with a regular lattice that avoids the fabrication constraints associated with prescribed concentric shells.

Aperture-free targets offer an additional reduction in fabrication complexity by eliminating the need to pattern a micrometre-scale opening. They also remove the requirement to align the laser with this channel, potentially simplifying target repositioning between destructive shots. These advantages may facilitate proof-of-principle experiments on solid-state LWFA, including those being prepared within the NanoAc collaboration, and offer a simpler option for applications that can tolerate the lower QMEP energies and increased divergence. Among the investigated configurations, $O_1$ provides an alternative when higher peak energy and lower divergence are priorities. In future work, spectral and angular beam-quality metrics will be incorporated into the objective function, and the laser parameters will be optimized jointly with the target geometry.

\begin{acknowledgments}
The authors acknowledge the National Laboratory for Scientific Computing (LNCC/MCTI, Brazil) for providing access to the Santos Dumont supercomputer (Project LPA-FARMA), which contributed to the research results reported in this paper. This work was supported by the Conselho Nacional de Desenvolvimento Científico e Tecnológico (CNPq) under Grant 405143/2021-4 and Scholarship 140941/2023-1, by Fundación Carolina, Estancias Cortas Postdoctorales 2026 (BEC-26), and by the Fundação de Amparo à Pesquisa do Estado do Rio Grande do Sul (FAPERGS) under Grant 24/2551-0001552-3.

\end{acknowledgments}

\section*{Data Availability Statement}
The data that support the findings of this study are available from the corresponding authors upon reasonable request.

\bibliographystyle{aipnum4-1-etal}
\bibliography{aipsamp}

\end{document}